\documentclass[twocolumn,showpacs,superscriptaddress,pra]{revtex4}
\usepackage{graphicx}
\usepackage{dcolumn}
\usepackage{bm}

\begin{document}

\title{Analogue of Cavity  QED  for Coupling between  Spin and Nanomechanical Resonator }

\author{Fei \surname{Xue}}
\email{Echoxue@itp.ac.cn}
\author{Ling \surname{Zhong}}
\author{Yong \surname{Li}}
\author{C.P. \surname{Sun}}
\email{suncp@itp.ac.cn} \ \homepage{http://www.itp.ac.cn/~suncp}
\affiliation{Institute of Theoretical Physics, Chinese Academy of
Sciences, 100080, China}

\date{\today }

\begin{abstract}
We describe a cavity QED analogue for the coupling system of a spin
and a nanomechanical resonator with a magnetic tip. For the
quantized nanomechanical resonator, a spin-boson model for this
coupling system can refer to a Jaynes-Cummings(JC) or an anti-JC
model. These observations predict some quantum optical phenomena,
such as squeezing and ``collapse-revival" in the single oscillation
mode of the nanomechanical resonator when it is initially prepared
in the quasi-classical state. By modulating the phase of RF magnetic
field one can switch the system between the JC and anti-JC model,
which provides a potential protocol for the detection of the single
spin. A damping mechanism is also analyzed.
\end{abstract}


\maketitle

\section{Introduction}
Many recent experiments have exhibited the possibilities to reach
GHz mechanical oscillations by a mechanical resonator in the
nano-scale\cite{Huang2003,Cleland2004,Gaidarzhy2005,Gaidarzhy2005b}.
With such a high oscillating frequency and at the temperature of mK,
the nanomechanical resonator(NAMR) should be modelled as a quantized
harmonic oscillator rather than a classical harmonic oscillator.
Such a quantized NAMR offers a realistic system to explore the new
phenomena in the quantum-classical crossover, such as the
quantization and the decoherence of generic macroscopic objects. And
the quantized nanomechanical resonator is a crucial component of the
nanomechanical analog of a laser \cite{Bargatin2003} or
maser\cite{Sun2006}. The experimental progresses were made in
nanomechanical oscillators, as well as in magnetic resonance force
microscopy(MFRM). Gaidarzhy \textit{et al.} demonstrated
GHz-frequency oscillations by a nano-scale mechanical resonator with
antenna structure \cite{Gaidarzhy2005}, while Rugar \textit{et al.}
demonstrated a single electron spin detection by the MRFM
\cite{Rugar2004}. In the MRFM, the system of spins and the NAMR is
often treated in a quasi-classical way \cite{Berman2005}, where the
NAMR is considered as a classical harmonic oscillator.

As the number of spins in the sample approaches one and the NAMR
reaches quantum regime, a fully quantum model is expected to
describe this coupling system correctly. In this paper we establish
a spin-boson model for this coupling system of the spin and the
NAMR. By quantizing the oscillation of the NAMR, a spin-boson model
which is an analog of the Jaynes-Cummings(JC) model in the cavity
QED and quantum optics, can be established for some setup
parameters. The successful implementation of such setup in the near
future experiments will lead to the so called mechanical QED
structure.

Such artificial engineered cavity QED structure can also demonstrate
the feature of the quantization of the NAMR through the physical
effects similar to various quantum optics phenomena. For example, in
the large detuning limit that the NAMR does not exactly resonate to
the spin, the virtual transition of the spin will result in the
interesting squeezing effect for the NAMR. In the resonant case, the
quantum dynamics of the coupling system is described by JC and
anti-JC models and thus the typical ``collapse-revival'' phenomenon
can be exhibited by the photon mode of the NAMR.

The paper is arranged as follows. In Sec.\ref{sec:model} a
spin-boson model is presented for the coupling system of the spin
and the NAMR. We then show in Sec.\ref{sec:squeezing} that the
single mode oscillation of the NAMR is dynamically squeezed in two
different ways with respect to the initial states of the spin $\vert
0 \rangle$ and $\vert 1 \rangle$ in the large detuning limit. In
Sec.\ref{sec:jc-anti-jc}, JC and anti-JC models are established for
the coupling system for the special phases of rotating RF
magnetic field, which suggests a potential way to detect single spin. In Sec.%
\ref{sec:damping} we consider the damping of the oscillation of the NAMR
that is exposed to thermal noise. In the end, there are some remarks in Sec.%
\ref{sec:remarks}.

\section{Spin-Boson model for coupling system of spin and nanomechanical
resonator}

\label{sec:model}

\begin{figure}[tp]
\centering
\includegraphics[scale=1]{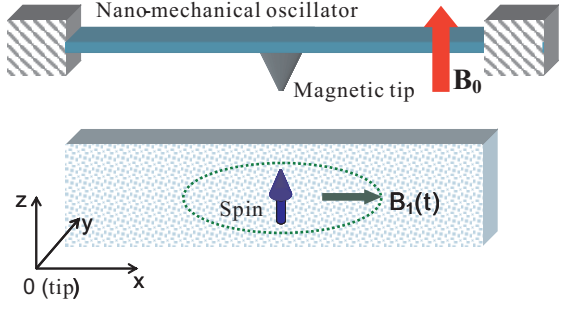}
\caption{(Color online) The setup of the spin and the nanomechanical
resonator. They are coupled with each other by a tiny ferromagnetic
particle attached on the NAMR.} \label{fig:fig-MQED}
\end{figure}

The schematics of the coupling system of the spin and the NAMR is
illustrated in Fig.\ref{fig:fig-MQED}. It is similar to that of the
MRFM, but here the NAMR is quantized, since it is assumed to
oscillate at GHz under a temperature of mK. A ferromagnetic particle
is glued on the middle of the NAMR and exerts a gradient magnetic
field on the spin below. Besides, there are also a static magnetic
field $B_0$ pointing in the $z$-direction and a rotating RF magnetic
field $B_1(t)$ imposed on the spin.

The magnetic tip produces a dipole magnetic field at the position of the
spin \cite{Jackson1999}:
\begin{eqnarray}
\vec{B}_{tip}=\frac{\mu _{0}}{4\pi }\frac{3(\vec{n}\cdot
\vec{m})\vec{n}- \vec{m}}{r^{3}},
\end{eqnarray}
where $\mu _{0}$ is the vacuum magnetic conductance, $\vec{n}$ the
unit vector from the tip to the spin, $\vec{m}$ the magnetic moment
of the ferromagnetic particle pointing in the $z$-direction, $r$ the
distance between the tip and the spin. The reference frame is
established with the origin point at the balance position of the
magnetic tip. For the sake of the simplicity we assume that the spin
is exactly beneath the tip and the NAMR with the magnetic tip
oscillates in the $x-z$ plane. The coordinate of the spin is set to
be $(0,0,d)$. As the vibration amplitude of the NAMR (the magnetic
tip) is very small comparing to the distance between the tip and the
spin, the magnetic field at the position of the spin produced by the
magnetic tip is approximately written as
\begin{eqnarray}
\vec{B}_{tip}(z)=(A-Gz)\hat{e}_{z}
\end{eqnarray}%
to the first order of approximation. Here, $z$ is the small
deviation of the magnetic tip from its balance position;
$\hat{e}_{k}$ ($k\in \{x,y,z\}$) is the unit vector for the $x,y,z$
axis respectively; $A=\mu _{0}m/(2\pi d^{3})$ and $G=3\mu
_{0}m/(2\pi d^{4})$.

The NAMR with the magnetic tip is modelled as a harmonic oscillator
in the lowest order of approximation, which is quantized in the
following. The Hamiltonian of the coupling system reads
\begin{eqnarray}
H=\frac{p_{z}^{2}}{2m_{eff}}+\frac{1}{2}m_{eff}\omega _{c}^{2}z^{2}+\gamma
\hbar \vec{B}\cdot \vec{S},
\end{eqnarray}%
where $m_{eff}$ is the effective mass of the harmonic oscillator,
$p_z$ the canonical momentum, $\omega _{c}$ its frequency, $-\gamma
$ the gyromagnetic ratio and
\begin{eqnarray}
\vec{B}=B_{0}\hat{e}_{z}+\vec{B}_{1}(t)+\vec{B}_{tip}(z).
\end{eqnarray}%
is the \ total magnetic field $.$The RF magnetic field $\vec{B}_{1}(t)$ is
\begin{eqnarray}
\vec{B}_{1}(t)=B_{1}\cos (\omega _{r}t+\varphi )\hat{e}_{x}+B_{1}\sin
(\omega _{r}t+\varphi )\hat{e}_{y}
\end{eqnarray}%
with $B_{1}$ being its amplitude, $\omega _{r}$ its frequency and $\varphi $
its phase. By performing a \textquotedblleft rotating reference frame"
transformation in the $x$-$y$ plane, the time dependence of $\vec{B}_{1}(t)$
can be eliminated. Then in the rotating frame  rotating  around the $z$%
-direction at the frequency $\omega _{r}$, the effective RF magnetic field
reads
\begin{eqnarray}
B_{1}^{\prime }=(\omega _{r}/\gamma )\hat{e}_{z}+B_{1}(\cos \varphi \hat{e}%
_{x}+\sin \varphi \hat{e}_{y}).
\end{eqnarray}%
By defining the bosonic operators $a^{\dagger }$ and $a$, the
momentum and coordinate operators $p_{z}$ and $z$ can be expressed
as their linear combinations:
\begin{eqnarray}
z=\sqrt{\frac{\hbar }{2m_{eff}\omega _{c}}}(a^{\dagger }+a),p_{z}=i\sqrt{%
\frac{m_{eff}\omega _{c}\hbar }{2}}(a^{\dagger }-a).
\end{eqnarray}

In the rotating reference frame the Hamiltonian of the coupling system can
be rewritten as $H=H_{B}+H_{S}+H_{I}$, where
\begin{eqnarray}
H_{B}=(a^{\dagger }a+\frac{1}{2})\hbar \omega _{c}
\end{eqnarray}%
is the Hamiltonian of the NAMR,
\begin{equation}
H_{S}=\gamma \hbar \lbrack B_{1}(\cos \varphi \tilde{S}_{x}+\sin \varphi
\tilde{S}_{y})+(B_{0}+A+\frac{\omega _{r}}{\gamma })\tilde{S}_{z}]
\end{equation}%
is the Hamiltonian of the spin part and
\begin{eqnarray}
H_{I}=-\gamma \hbar \lambda G\tilde{S}_{z}(a^{\dagger }+a)
\end{eqnarray}%
describes the interaction between the spin and the NAMR. Here, $\tilde{S}%
_{k}=(1/2)\sigma _{k}$ ($k\in \{x,y,z\}$) are the spin operators and $%
\lambda \equiv \sqrt{{\hbar }/({2m_{eff}\omega _{c}})}$.

We set
\begin{equation}
B_{0}+A+\frac{{\omega _{r}}}{{\gamma }}=0
\end{equation}
with the special setup parameters, i.e., the frequency of the RF
magnetic
field is just at the resonant point. By rotating the axes of spin as: $%
S_{z}\equiv \tilde{S}_{x}$, $S_{y}\equiv \tilde{S}_{y}$ and $S_{x}\equiv -%
\tilde{S}_{z}$, we obtain the following Hamiltonian for the coupling system:
\begin{eqnarray}
H &=&\hbar \omega _{c}(a^{\dagger }a+\frac{1}{2})+\gamma \hbar B_{1}(\cos
\varphi S_{z}+\sin \varphi S_{y})  \nonumber \\
&&+\hbar g(a^{\dagger }S_{-}+aS_{+}+aS_{-}+a^{\dagger }S_{+}),
\label{eqn:Main-Hamiltonian}
\end{eqnarray}%
where $a^{\dagger }$ and $a$ are creation and annihilation operators
for the NAMR, $S_{\pm }\equiv S_{x}\pm iS_{y}$ are raising and
lowering operators for the spin respectively, and the coupling
constant between the mechanical oscillator and the spin is $g\equiv
\gamma \lambda G/2$. In the following sections we will explore two
quantum features of this spin-boson model in two cases: on-resonance
and off-resonance situation.

\section{Dynamically squeezing of the oscillation of the nanomechanical
resonator}

\label{sec:squeezing}

In the case of off-resonance situation, there is no obvious exchange
of energy between the spin and NAMR, but the squeezing effects of
the oscillations of the NAMR appear because of the interference of
the two oscillation modes with respect to the two eigenstates of the
spin. As is well known, the squeezed state is one kind of
minimum-uncertainty state. The quantum fluctuation of one quadrature
component of the squeezed state is less than that of the coherent
state. Thus the uncertainty of the position or the momentum of the
squeezed NAMR may be reduced below that in standard quantum limit
\cite{SQL}. In the following we will show that the oscillation of
the NAMR can be squeezed by the off-resonance interaction with the
spin. Similar quantum squeezing of mechanical motion for the NAMR is
discussed by Blencowe and Wybourne with the capacitive coupling NAMR
and its substrate \cite{Blencowe2000}, and by Wang \textit{et al.}
with the NAMR coupling to a Cooper pair box \cite{WYD2004}.

The off-resonance interaction, i.e., in large detuning situation
$g\ll \Delta \equiv |\omega _{s}-\omega _{c}|$, Hamiltonian
(\ref{eqn:Main-Hamiltonian}) can be approximately written in a
diagonal form by adiabatically eliminating coherent effect between
the spin states $|0\rangle $ and $|1\rangle $ \cite{WYD2004} :
\begin{eqnarray}
H_{diag}=\sum_{k=0,1}H_{k}\otimes |k\rangle \langle k|,
\end{eqnarray}%
where we have put $\varphi =0$ in Hamiltonian (\ref{eqn:Main-Hamiltonian}).
The explicit form of $H_{k}$ is
\begin{eqnarray}
H_{k} &\approx &\omega _{c}a^{\dagger }a+\frac{(-1)^{k}}{2}\omega _{s}
\nonumber \\
&&-\frac{(-1)^{k}g^{2}}{4\Delta }(a^{\dagger }a^{\dagger }+2a^{\dagger
}a+aa+1),
\end{eqnarray}%
where $\omega _{s}=\gamma B_{1}$.

To diagonalize each effective Hamiltonian $H_{k}$, we use the Bogliubov
transformation or the called squeezing transformation:
\begin{equation}
b_{k}=\mu _{k}a-\nu _{k}a^{\dagger },
\end{equation}%
where the parameters
\begin{eqnarray}
\mu _{k} &=&\frac{1}{2}\left( \sqrt{N_{k}}+\frac{1}{\sqrt{N_{k}}}\right) ,
\nonumber \\
\nu _{k} &=&\frac{1}{2}\left( \sqrt{N_{k}}-\frac{1}{\sqrt{N_{k}}}\right)
\end{eqnarray}%
are defined in terms of
\begin{equation}
N_{k}=\sqrt{\frac{\omega _{c}\Delta }{\omega _{c}\Delta -(-1)^{k}g^{2}}}.
\end{equation}%
The diagonal Hamiltonian then reads
\begin{equation}
H_{k}=\Omega _{k}(b_{k}b_{k}^{\dagger }+\frac{1}{2})+E_{k}(-1)^{k}(\frac{%
\omega _{s}}{2}-\frac{g^{2}}{4\Delta }),
\end{equation}%
where $E_{k}(0)=(-1)^{k}(\omega _{s}/2-g^{2}/(4\Delta )).$ The
eigen-frequency
\begin{equation}
\Omega _{k}\approx \omega _{c}\left( 1-(-1)^{k}\frac{g^{2}}{2\omega
_{c}\Delta }\right) .
\end{equation}%
contains a term $\delta =g^{2}/2\omega _{c}\Delta ,$ which is called AC
Stark shift and is essentially due to the Lamb effect.

Since the coherent state of the NAMR is most close to the classical one, it
seems to be feasible to prepare the NAMR in this state. Starting from the
coherent state $\left\vert \alpha \right\rangle $, at time $t$ the NAMR will
evolve into
\begin{equation}
\left\vert \Psi \left( t\right) \right\rangle _{k}=e^{-i\Omega
_{k}b_{k}^{\dag }b_{k}t}\left\vert \alpha \right\rangle
\end{equation}%
with respect to the spin state $|k\rangle $ according to the above
discussion.

To evaluate the squeezing property of $\left\vert \Psi \left( t\right)
\right\rangle _{k}$, we first consider $B_{k}\left( t\right) \left\vert \Psi
\left( t\right) \right\rangle _{k}$, where
\begin{equation}
B_{k}\left( t\right) =\mu _{k}(t)b_{k}+\nu _{k}(t)b_{k}^{\dag }
\end{equation}%
is the time evolution of the Heisenberg operators $b_{k}$, and is found to
be dynamical squeezing operators. Here, the time-dependent coefficients are
\begin{equation}
\mu _{k}(t)=\mu _{k}e^{i\Omega _{k}t},\nu _{k}(t)=\nu _{k}e^{-i\Omega _{k}t}.
\end{equation}

\begin{figure}[tp]
\centering
\includegraphics[scale=1]{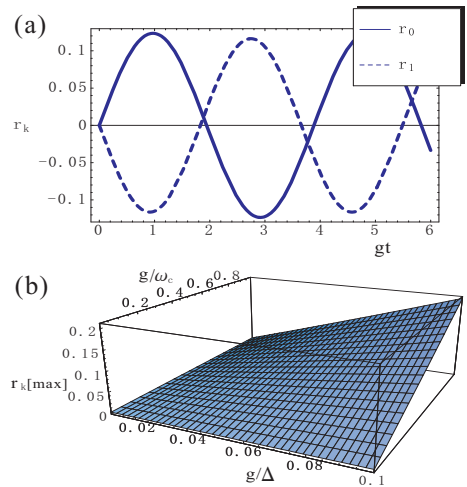}
\caption{(Color online) (a) The squeezing factor $r_k$ of the
oscillation of the NAMR initially in the coherent state $\vert
\protect\alpha \rangle$ for the different state $\vert k \rangle$ of
the spin. (b) The maximum squeezing factor $ln N_k$. The detuning
$g/\Delta \in [0.01,0.1]$ and coupling strength $g/ \protect\omega_c
\in [0.01,0.99]$.} \label{fig:fig-02ab}
\end{figure}

In fact, the state $\left\vert \Psi \left( t\right) \right\rangle _{k}$ of
the NAMR is the eigenstate of squeezing operator $B_{k}\left( t\right) $ and
the corresponding eigenvalue is $\alpha $:
\begin{equation}
B_{k}\left( t\right) \left\vert \Psi \left( t\right) \right\rangle
_{k}=\alpha \left\vert \Psi \left( t\right) \right\rangle _{k}.
\label{eqn:Bk-eigen}
\end{equation}%
Now we can write the quasi-excitations of $B_{k}(t)$ and $B_{k}(t)^{\dagger }
$ in terms of the elementary photon mode of $a$ and $a^{\dagger }$:
\begin{equation}
B_{k}\left( t\right) =u_{1}\left( t\right) a+u_{2}\left( t\right) a^{\dagger
},
\end{equation}%
where
\begin{eqnarray}
u_{1}\left( t\right)  &=&\cos \Omega _{k}t+i\left( \mu _{k}^{2}+\nu
_{k}^{2}\right) \sin \Omega _{k}t, \\
u_{2}\left( t\right)  &=&2i\mu _{k}\nu _{k}\sin \Omega _{k}t,
\end{eqnarray}%
and
\begin{equation}
\left\vert u_{1}\left( t\right) \right\vert ^{2}-\left\vert u_{2}\left(
t\right) \right\vert ^{2}=1.
\end{equation}%
According to Eq.(\ref{eqn:Bk-eigen}), the time evolution of the NAMR in the
case of off-resonance follows the eigenstate of the quasi-excitation of $%
B_{k}(t)$. And the eigenstate of $B_{k}\left( t\right) $ is found to be
squeezed state with respect to the elementary excitation of the NAMR of $a$
and $a^{\dagger }$. The squeeze factor is calculated as
\begin{equation}
r_{k}\left( t\right) =\ln \left( 2\mu _{k}\nu _{k}\sin \Omega _{k}t+\sqrt{%
1+4\mu _{k}^{2}\nu _{k}^{2}\sin ^{2}\Omega _{k}t}\right) .
\end{equation}

The dynamical squeezings of the oscillations of the NAMR are plotted in Fig.%
\ref{fig:fig-02ab} (a), in which the time is in the unit of $1/g$. Depending
on the spin state $|k\rangle $, the NAMR is dynamically squeezed with
different frequencies $\Omega _{k}$ and with also the different maximum
squeeze factor $\ln N_{k}$. The maximum squeezing $\ln N_{k}$ of the
oscillation of the NAMR depends on the detuning $g/\Delta $ and the strength
of coupling $g/min\{\omega _{c},\omega _{s}\}$, which is shown in Fig.\ref%
{fig:fig-02ab} (b). Here, we have assumed that $\omega _{c}<\omega
_{s}$ in the plotting for convenience. The greater squeezing occurs
with smaller detuning and stronger coupling. To guarantee the large
detuning assumption in the above discussions, the detuning can not
be too small. However, large squeezing can still be attained with
the appropriate coupling strength.

Similar to the JC model, in large detuning situation this spin-boson system
exhibits \textquotedblleft collapse-revival" phenomenon. When the NAMR is
initially prepared in the coherent state $\left\vert \alpha \right\rangle $,
its position $\langle x\rangle _{t}$ can be calculated by noticing the fact
that the time evolution of the NAMR always follows the eigenstate of
quasi-excitation $B_{k}\left( t\right) $. Therefore, we expand $a+a^{\dag }$
in terms of $B_{k}\left( t\right) $ and $B_{k}^{\dag }\left( t\right) $,
then $\left\langle x\right\rangle _{t}=\left\langle a+a^{\dag }\right\rangle
_{t}$ is calculated as
\begin{eqnarray}
\left\langle x\right\rangle _{t} &=&\sum_{k}c_{k}c_{k}^{\ast }(\alpha \left[
\frac{1}{N_{k}}\cos \Omega _{k}t-i\sin \Omega _{k}t\right]   \nonumber \\
&&+\alpha ^{\ast }\left[ \frac{1}{N_{k}}\cos \Omega _{k}t+i\sin \Omega _{k}t%
\right] ),
\end{eqnarray}%
where the spin is prepared in the state $c_{0}|0\rangle +c_{1}|1\rangle $
initially. When the spin is prepared in the superposition states, the
oscillation of the NAMR collapses and revives, as illustrated in Fig.\ref%
{fig:fig-03}. This is the result of the interference effect of the
two oscillating mode $\Omega _{k}$ labelled by the spin state
$|k\rangle $, which is a witness of the quantization of the NAMR.

\begin{figure}[tp]
\centering
\includegraphics[scale=0.6]{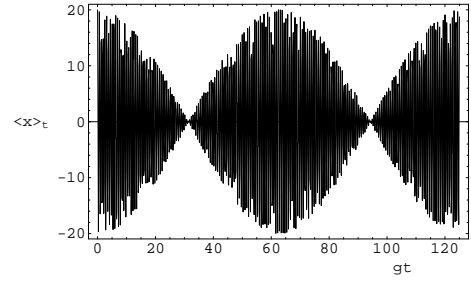}
\caption{Oscillating of the NAMR under off-resonance situation for the
system of the NAMR and the spin. }
\label{fig:fig-03}
\end{figure}

\section{Switching between JC Hamiltonian and anti-JC Hamiltonian}

\label{sec:jc-anti-jc}

When the spin resonate with the NAMR, under the rotating wave
approximation, one can obtain the JC model or anti-JC model from the
spin-boson model of Eq.(\ref{eqn:Main-Hamiltonian}). Then the spin
and the NAMR form a ``mechanical QED'' structure -- an analog of the
cavity QED. It is exciting that the ``mechanical QED'' offers a new
analogue of cavity QED
among various solid state cavity QED systems \cite%
{Armour2002,Irish2003,Blais2004,Wallraff2004,Chiorescu2004,Wang2004} .

Without loss of generality, in the following we assume that $B_1 > 0$. When
the phase of the RF magnetic field $\varphi =0$, the Hamiltonian (\ref%
{eqn:Main-Hamiltonian}) becomes a JC Hamiltonian
\begin{eqnarray} \nonumber
H_{JC} &=& \hbar \omega_c (a^{\dagger}a + \frac{1}{2}) + \hbar
\omega_s S_z + \hbar g(a^{\dagger}S_- + a S_+ ),  \\
\label{eqn:JC-Hamiltonian}
\end{eqnarray}
under the rotating wave approximation. Here $\omega_s = \gamma B_1 > 0$.
This can be understood in the interaction picture, in which the term
\begin{eqnarray}
F_1=e^{-i(\omega_c + \omega_s)t/\hbar}a S_- + e^{i(\omega_c +
\omega_s)t/\hbar} a^{\dagger}S_+
\end{eqnarray}
oscillates fast with higher frequency, and thus can be neglected. For this
JC Hamiltonian in the large detuning limit $g/\Delta \ll 1$, where $%
\Delta=|\omega_c-\omega_s|$, the dressed energy level of the NAMR is $%
\omega_c + {2g^2} \langle S \rangle /{\Delta} $.

When the phase of the RF magnetic field $\varphi =\pi$, the Hamiltonian (\ref%
{eqn:Main-Hamiltonian}) becomes an anti-JC Hamiltonian
\begin{eqnarray} \nonumber
H_{AJC} = \hbar \omega_c (a^{\dagger}a + \frac{1}{2}) + \hbar
\omega_s S_z + \hbar g( aS_- + a^{\dagger}S_+ ), \\
\label{eqn:anti-JC-Hamiltonian}
\end{eqnarray}
under the rotating wave approximation. Here $\omega_s = -\gamma B_1 < 0$.
This can be understood in the interaction picture, in which the term
\begin{eqnarray}
F_2=e^{-i(\omega_c - \omega_s)t/\hbar} aS_+ + e^{i(\omega_c -
\omega_s)t/\hbar} a^{\dagger} S_-,
\end{eqnarray}
rather than $F_1$, oscillates fast with higher frequency, thus
should be neglected. For this anti-JC Hamiltonian at the large
detuning situation, the dressed energy level of the NAMR reads
$\omega_c - {2g^2} \langle S \rangle/{\Delta} $.

\begin{figure}[tp]
\centering
\includegraphics[scale=0.8]{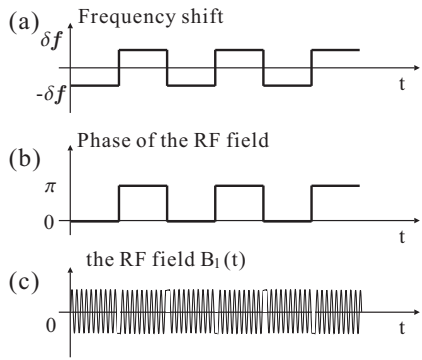}
\caption{Timing diagram for switching the coupling system between
the JC and the anti-JC Hamiltonian. The initial state of the spin is
supposed to be the ground state of the JC Hamiltonian. The frequency
shift $\protect\delta f = {g^2}/{\Delta}$ of the NAMR changes its
sign when the system switches between the two Hamiltonians, while
the state of the spin evolves little.} \label{fig:fig-pulses}
\end{figure}

In the two situations of the JC and the anti-JC models that we
discussed above, it is noticed that the AC Stark shifts have
opposite signs. This motivates us to present a scheme for the single
spin detection. The trick is similar to the interrupted oscillating
the cantilever-driven adiabatic reversals (iOSCAR) protocol
\cite{Rugar2004}, and provide a convenient frequency shift detection
without requesting the difficult measurement of the absolute
frequency of the NAMR. The RF magnetic field is interrupted for half
a cycle periodically as illustrated in Fig.\ref{fig:fig-pulses} (c).
Then the phase of the RF magnetic field $\varphi$ will swing between
$0$ and $\pi$ correspondingly, which is illustrated in
Fig.\ref{fig:fig-pulses} (b). Therefore, as discussed above, the
``mechanical QED'' of the NAMR and the spin switches between the JC
Hamiltonian and the anti-JC Hamiltonian. When the switching is fast
enough so that the spin only evolves little and is kept in its
original state, i.e., the expectation of the spin $\langle S
\rangle$ keeps its value, the frequency shift of the NAMR swings
between ${2g^2} \langle S \rangle /{\Delta}$ and $- {2g^2}\langle S
\rangle /{\Delta}$ periodically as the consequence of this
Hamiltonian switching, which is illustrated in
Fig.\ref{fig:fig-pulses} (a).

To discuss the frequency shifts of the NAMR with concrete parameters
in practical experiments, we write down the expression of the
coupling constant
\begin{eqnarray}
g = \gamma G\sqrt{\frac{\hbar}{8m_{eff}\omega_c}} \approx 10^{-7} G \sqrt{%
\frac{\omega_c}{k_{eff}}}  \label{eqn:g-factor}
\end{eqnarray}
explicitly, here $k_{eff}$ is the effective spring constant of the
mechanical oscillator. Substituting the parameters of
Ref.\cite{Rugar2004}, where $G \sim 10^5$ T/m, $k_{eff}=0.11$ mN/m
and $\omega_c=5.5$ kHz, in the above equation, we obtain a frequency
shift about 7 Hz for the NAMR. Here we have supposed the detuning
$g/\Delta=0.1$. The frequency shift is larger than that in the
current OSCAR technique (about 4 mHz in Ref.\cite{Rugar2004}) by
three orders of magnitude. In the OSCAR the NAMR is treated as a
classical oscillator. Therefore, if the temperature of the system
can be reduced to $\hbar \omega_c / k_B = 42$ nK so that it enters
the quantum regime, mechanical QED model indeed provides an
efficient protocol for single electron spin detection.

There are some protocols \cite{Hopkins2003,Hohberger2004, Zhang2005}
that promise to cool the temperature of the NAMR down to mK or even
lower. However reducing the temperature of the NAMR still remains a
great challenge in the experiments. To our best knowledge, the
lowest temperature attained in the dilution fridge now is about 30
mK. So we consider another way to push the NAMR into the quantum
regime by increasing its frequency rather than decreasing its
temperature. Unfortunately, the effective spring constant of the
simple structure resonator increases rapidly with the oscillator
frequency: $k_{eff} \propto \omega_c^2$. When the
frequency of the NAMR increases, the coupling constant $g \propto 1/\sqrt{%
\omega_c}$ decreases. Therefore a much smaller frequency shift about
0.01 Hz occurs for a 1.5 GHz nanomechanical oscillator with the
simple structure like the one in Ref.\cite{Rugar2004}, which makes
it hard to be detected in the experiments. However, in
Ref.\cite{Gaidarzhy2005}, an elaborately designed nanomechanical
resonator with an antenna structure reaches the frequency of 1.5 GHz
with a smaller effective spring constant about 300 N/m. Therefore,
for this enhanced resonator, the frequency shift is about 3 Hz ,
which is larger than that of the simple structure resonator by about
two orders of magnitude. As indicated in Eq.(\ref{eqn:g-factor})
stronger coupling constant $g$ can be reached by increasing the
magnetic field gradient and decreasing the effective spring constant
of the nanomechanical resonator. The stronger the coupling is, the
larger frequency shifts is obtained.

Developments in the experiments are still necessary to practise the
presented protocol in detecting single spin. However, as no
adiabatic condition is requested in this protocol, the frequency of
the NAMR can be set to resonant with the spin, while for the single
spin detection with the OSCAR technique, such as
Ref.\cite{Rugar2004,Berman2005}, the frequency of the NAMR must be
much smaller than that of the spin. So if one sets the frequency of
the NAMR at MHz, then a micro Kelvin temperature is sufficient to
prepare the NAMR in the ground state, which is much more realizable
in the experiments in the near future. And the frequency shift to be
detected is about tens Hz. The required resolving power is about
$10^{-6}$. It is the same to that of the OSCAR MRFM mHz/kHz=
$10^{-6}$ \cite{Rugar2004}, which has already been achieved in the
experiments. There is indeed no fundamental difficulty of the above
suggested single spin detection. Actually there are still space to
further optimize the frequency of the NAMR to balance the resolving
power and the temperature to be attained. The discussion in this
section suggests a single spin detection in the absence of the
environment, which is far from the reality. In the next section we
will discuss the influence of the thermal environments on the NAMR.

\section{Damping of the nanomechanical resonator}

\label{sec:damping}

\begin{figure}[tp]
\centering
\includegraphics[scale=0.7]{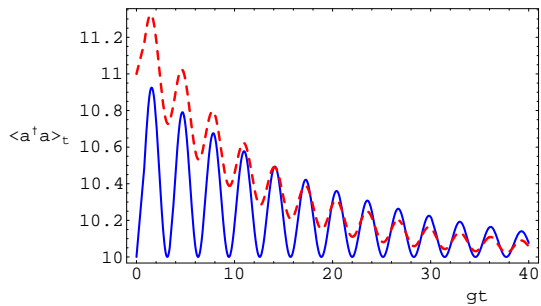}
\caption{(Color on line) The damping of the NAMR. The energy of the
NAMR $\langle a^{\dagger}a \rangle_t$ is plotted with the NAMR or
spin in excited state under weak coupling environment situation
$\protect\kappa/g=0.2$. The red(dashing) line is for the NAMR
initially in the excited state, and the blue(solid) line is for the
spin initially in the excited state.} \label{fig:fig04}
\end{figure}

In the above discussions, we demonstrate various quantum optical
phenomena due to the quantization of the NAMR. But due to the
coupling to the complex environments in practice, these phenomena
may be washed out. Therefore it is necessary to study the influence
of environments. Both the spin and the NAMR are coupled to the
environments, which causes the  system damp their energy into the
environments in thermal equilibrium. It is suggested by the
experiments of the MRFM that the spin has longer relaxation time
than the NMR \cite{Rugar2004}. For the detailed  discussions about
the
dissipation in the NAMR, one can refer to Ref. \cite%
{Mohanty2002,Berman2003,Gassmann2004}. The damping behavior of the total
system is mainly dominated by that of the NAMR. In general, we model the
environment as a multi-mode boson bath coupling to the NAMR. The total
Hamiltonian of the coupling system of the NAMR and the spin and the
environment reads
\begin{eqnarray}
H_{total}=H+\sum_{j}\hbar \omega _{j}b_{j}^{\dagger }b_{j}+\hbar
\sum_{j}k_{j}(a^{\dagger }b_{j}+ab_{j}^{\dagger }).
\end{eqnarray}%
Here , $\ H$ is the Hamiltonian (\ref{eqn:Main-Hamiltonian}) of the
total system, $b_{j}$ and $b_{j}^{\dagger }$ are creation and
annihilation operators of the $j$th mode of the environment, $k_{j}$
is the coupling constant between the NAMR and the $j$th mode of the
environment. We neglect various two-excitation processes, and terms
such as $ab_{j}$ and $a^{\dagger }b_{j}^{\dagger }$ are omitted in
the total Hamiltonian.

We invoke Heisenberg-Langevin method to deal with this quantum
damping problem. The operators $(a^{\dagger })^{m}a^{n}O$,
($O=S_{z},S_{\pm }$), satisfy the following equations: \cite{Scully}
\begin{eqnarray}
&&\frac{d}{dt}[(a^{\dagger })^{m}a^{n}O]=-\frac{i}{\hbar }[(a^{\dagger
})^{m}a^{n}O,H]  \nonumber \\
&&+(i\kappa Q(m-n)-\frac{1}{2}\kappa (m+n))\langle (a^{\dagger
})^{m}a^{n}\rangle _{E}  \nonumber \\
&&+\kappa mnn_{th}\langle (a^{\dagger })^{m-1}a^{n-1}\rangle _{E},
\end{eqnarray}%
where $\kappa $ is the damping factor, $Q$ the quality factor of the
NAMR, and $\langle ...\rangle _{E}$ represents tracing of the
environment.

The higher order processes, such as the terms $ab_{j}$ and
$a^{\dagger }b_{j}^{\dagger }$, of the NAMR are neglected in the
following discussions. Therefore the expectations of the operators
including quadratic or higher powers in operators $a$ and
$a^{\dagger }$ of the NAMR, such as $\langle a^{2}\rangle $ and
$\langle a^{\dagger 2}S_{z}a\rangle $, are zero. With these
approximations we simultaneously obtain the closed system of
equations:
\begin{eqnarray}
\frac{d\langle a^{\dagger }a\rangle }{dt} &=&-\kappa \langle a^{\dagger
}a\rangle +gX+\kappa n_{th} \\
\frac{d\langle S_{z}\rangle }{dt} &=&-2gX \\
\frac{dX}{dt} &=&g\langle S_{z}\rangle -\frac{1}{2}\kappa X+2g\langle
a^{\dagger }S_{z}a\rangle +g \\
\frac{d\langle a^{\dagger }S_{z}a\rangle }{dt} &=&-gX-\kappa \langle
a^{\dagger }S_{z}a\rangle
\end{eqnarray}%
where $n_{th}$ is the average thermal photons in the environment,
and $X=i\langle S^{+}a+S^{-}a^{\dagger }\rangle$.  The above system
of equations can be solved by Laplace transform method or
numerically. In Fig.\ref{fig:fig04} under situation of weak coupling
to the environment, we plot the time evolving of the expectation of
the energy of the NAMR $\langle a^{\dagger }a\rangle $ with the
initial condition $n_{th}=10$, $\langle a^{\dagger }a\rangle
+\langle S_{z}\rangle =11$, $X=\langle a^{\dagger }S_{z}a\rangle
=0$. The case of stronger coupling to the environment, i.e., larger
$\kappa /g$, is also studied. It is not surprising that the
oscillation of the energy of the NAMR die out faster with larger
$\kappa /g$, and with sufficient strong coupling to environment, no
oscillation of the energy of the NAMR could be observed. This simple
consideration suggests that even for non-zero temperature it is
still possible to observe the energy oscillation between the NAMR
and the spin as long as the coupling to the environment is weak
enough.

\section{Summary and Remarks}

\label{sec:remarks}

We have revisited a spin-boson model for the coupling system of the
nanomechanical resonator and the spin. With the same blocks (the
NAMR with a mgnetic tip and the spin) as that in the MRFM, but with
the refined setup parameters, the NAMR enters the quantum regime and
thus a fully quantum theory is needed. With the quasi-classical
state (the coherent state) as the initial state of the NAMR, the
dynamical squeezing of the mode and also the oscillation of the NAMR
is studied under the large detuning limit. Both the coherent state
and the low frequency (large detuning) of the macro-size resonator
are preferred in the experiments. Squeezed states, even if there is
the thermal noise and the displacement, can have non-classical
properties, such as bunch distribution of phonon
number\cite{Marian1993a,Marian1993b,Lu2000}. With the model setup
and the predications of the quantum optical phenomena, we actually
describe a cavity QED analog for the quantum dynamics of the
coupling system. For the quantized NAMR, the spin-boson model for
this coupling system can refer to a JC or an anti-JC model according
to different physical accessible parameters. By modulating the phase
of RF magnetic field one can switch the system between the JC and
anti-JC model. This observation provides a potential protocol for
the detection of the single spin.

Acknowledgments: This work is supported by the NSFC with grant Nos.
90203018, 10474104, 60433050 and 10574133. It is also funded by the
National Fundamental Research Program of China with Nos.
2001CB309310 and 2005CB724508. We also thank Y.D. Wang and Yong Li
for helpful discussions.


\begin{thebibliography}{99}
\bibitem{Huang2003} X.M.H. Huang, C.A. Zorman, M. Mehregany and ML Roukes,
Nature (London) \textbf{421}, p496 (2003).

\bibitem{Cleland2004} A. N. Cleland and M. R. Geller, Phys. Rev. Lett.
\textbf{93}, 070501, (2004).

\bibitem{Gaidarzhy2005} A. Gaidarzhy, G. Zolfagharkhani, R. L. Badzey and P.
Mohanty, Phys. Rev. Lett. \textbf{94}, 030402 (2005).

\bibitem{Gaidarzhy2005b} A. Gaidarzhy, G. Zolfagharkhani, R. L. Badzey and
P. Mohanty, Appl. Phys. Lett. \textbf{86}, 254103 (2005).

\bibitem{Bargatin2003} Igor Bargatin and M. L. Roukes, Phys. Rev. Lett.
\textbf{91} 138302 (2003).

\bibitem{Sun2006} C. P. Sun, L. F. Wei, Yu-xi Liu, and Franco Nori, Phys.
Rev. A \textbf{73}, 022318 (2006).

\bibitem{Rugar2004} D. Rugar, R. Budakian, H. J. Mamin and B. W. Chui,
Nature \textbf{430}, p329 (2004).

\bibitem{Berman2005} G. P. Berman, F. Borgonovi, V.N.Gorshkov and V. I.
Tsifrinovich, IEEE Transactions on Nanotechnology \textbf{4}, p14-20 (2005).

\bibitem{Jackson1999} John David Jackson, 1999, 3rd Ed., \textit{Classical
Electrodynamics}, p186, (John Willey \& Sons Inc. Press).

\bibitem{SQL} V. B. Braginsky, F.Y. Khalili, \textit{Quantum Measurement,}
Cambridge Univ Pr (June 1995).

\bibitem{Blencowe2000} M. P. Blencowe and M. N. Mybourne, Physica B(LT 22
Proceedings) \textbf{280}, p555, (2000).

\bibitem{WYD2004} Y. D. Wang, Y.B. Gao and C. P. Sun, Eur. Phys. J. B
\textbf{40}, p321-326, (2004).

\bibitem{Armour2002} A. D. Armour, M. P. Blencowe and K. C. Schwab, Phys.
Rev. Lett. \textbf{88}, 148301 (2002).

\bibitem{Irish2003} E. K. Irish and K. Schwab, Phys. Rev. B \textbf{68},
155311 (2003)

\bibitem{Blais2004} A. Blais, R. Huang, A. Wallraff, S. Girvin and R.
Schoelkopf, Phys. Rev. A \textbf{69}, 062320 (2004)

\bibitem{Wallraff2004} A. Wallraff, D. L. Schuster, A. Blais, L. Frunzio, R.
-S. Huang, J. Majer, S. M. Girvin and R. J. Schoelkopf, Natrue \textbf{431},
p162 (2004).

\bibitem{Chiorescu2004} I. Chiorescu, P. Bertet1, K. Semba, Y. Nakamura, C.
J. P. M. Harmans and J. E. Mooij, Nature \textbf{431}, p159 (2004).

\bibitem{Wang2004} Y. D. Wang, P. Zhang, D. L. Zhou, and C. P. Sun, Phys.
Rev. B \textbf{70}, 224515 (2004).

\bibitem{Hopkins2003} Asa Hopkins, Kurt Jacobs, Salman Habib and Keith
Schwab, Phys. Rev. B \textbf{68}, 235328 (2003).

\bibitem{Hohberger2004} Constanze H\"{o}hberger Metzger and Khaled Karral,
Nature \textbf{432}, p1002 (2004).

\bibitem{Zhang2005} P. Zhang, Y. D. Wang, and C. P. Sun, Phys. Rev. Lett.
\textbf{95}, 097204 (2005).

\bibitem{Mohanty2002} P. Mohanty, D. A. Harrington, K. L. Ekinci, Y. T.
Yang, M. J. Murphy and M. L. Roukes, Phys. Rev. B \textbf{66},
085416 (2002).

\bibitem{Berman2003} G. P. Berman, F. Borgonovi, Hsi-Sheng Goan, S. A.
Gurvitz, and V. I. Tsifrinovich, Phys. Rev. B \textbf{67}, 094425
(2003).

\bibitem{Gassmann2004} Hanno Gassmann, Mahn-Soo Choi, Hangmo Yi, and C.
Bruder, PRB \textbf{69}, 115419 (2004).

\bibitem{Scully} M.O. Scully, M.S Zubairy, \textit{Quantum Optics},
Cambridge University Press (January 1997)

\bibitem{Marian1993a} Paulina Marian and Tudor A. Marian, Phys. Rev. A
\textbf{47}, 4474 (1993)

\bibitem{Marian1993b} Paulina Marian and Tudor A. Marian, Phys. Rev. A
\textbf{47}, 4487 (1993)

\bibitem{Lu2000} W-F. Lu, J. Phys. A: Math. Gen. \textbf{33} 479 (2000).

\end{thebibliography}
\end{document}